\documentclass[11pt,twoside]{article}
\usepackage{epsfig}
\begin{document}

{\bf \noindent NEW FRONTIERS IN COSMOLOGY \\ AND  GALAXY FORMATION:\\
 CHALLENGES FOR THE FUTURE}
\footnote
{Proceeding of conference "Structure formation in the universe", held in Chamonix 2007. To be published in Structure formation in Astrophysics, Ed. G. Chabrier, Cambridge University Press, 2008}

\bigskip

{\noindent Richard Ellis$^1$ and Joseph Silk$^2$\\

\noindent $^1$Department of Astronomy, California Institute of Technology, Pasadena CA 91125, USA\\
$^2$Department of Physics, University of Oxford, Oxford OX1 3RH, UK}


\def\simlt{\lower.5ex\hbox{$\; \buildrel < \over \sim \;$}}
\def\simgt{\lower.5ex\hbox{$\; \buildrel > \over \sim \;$}}
\def\simpropto{\lower.2ex\hbox{$\; \buildrel \propto \over \sim \;$}}

\vspace{1truecm}

\centerline{\bf ABSTRACT}
\vspace{0.5truecm}
 
{ Cosmology faces three distinct challenges in the next decade. 
(1) The dark sector, both dark matter and dark energy, dominates the universe.  
Key questions include determining the nature of the dark matter and whether 
dark energy can be identified with, or if dynamical, replace,  the cosmological constant.  Nor, given the heated level of current debates about the nature of gravity and string theory, can one yet  unreservedly accept that dark matter or the cosmological constant/dark energy actually exists. Improved observational probes are crucial in this regard. (2) Galaxy formation was initiated at around the epoch of reionisation: we need to understand how and when the universe was reionised, as well as to develop probes of what happened at earlier epochs.  (3) Our simple dark matter--driven picture of galaxy assembly is seemingly at odds with several observational results, including the presence of ULIRGS at high redshift, the `downsizing' signature whereby massive objects terminate their star formation prior to those of lower masses, chemical signatures of $\alpha-$ element ratios in early-type galaxies,  and suggestions that merging may not be important in defining the Hubble
sequence. Any conclusions, however, are premature, given current uncertainties about possible hierarchy-inverting processes involved with feedback. Understanding the physical implications of these observational results in terms of a model of star formation in galaxies is a major challenge for theorists, and refining the observational uncertainties is a major goal for observers.}

\section{Introduction}

We live in interesting times where we can measure, with good and improving precision, the
constituents of our Universe. However, we should not confuse measurement
with understanding: a major puzzle is that only 4\% of the present energy
density of the cosmos is in familiar baryonic form. The first challenge we will
discuss in this brief review is how to move from measurement to understanding
in the most basic question of all - what constitutes our Universe?

The key, we will argue, is improved observations. Promising tools are 
available to tackle the distribution of dark matter on various scales, as well as to 
trace the history of the dark energy that causes the 
recently discovered 
cosmic acceleration. On a more fundamental level, the theoretical challenge of the cosmological constant and dark energy is so demanding in terms of fundamental physics that significant numbers of theorists are searching for 
alternative explanations of the observed acceleration. The existence of dark matter also is being questioned in schemes of alternative theories of gravitation.
Refinement of the observational probes is essential in order to consolidate both the nature of the acceleration which motivates dark energy and the  theoretical  infrastructure of Einstein/Newton gravity that provides the framework for dark matter.  Much investment will be needed to calibrate and
understand the limitations of these various approaches, and dedicated
missions will likely be needed, but the science returns will be considerable. Whatever its origin, dark energy promises to fundamentally change our views of the physical world. And dark matter already has brought fundamental change to our understanding of structure formation, although  more questions have been raised than answered.

Excellent observational progress is also being made in pushing the redshift frontiers. The most distant known galaxy, at the time of writing, lies at $z\simeq$7  and promising candidates are now being claimed to $z\simeq$10. Searches for  the earliest stellar systems are motivated by more than redshift records. Poorly-understood processes govern the distribution of luminosities and masses of the first stellar systems yet these objects acts as the basis from which galaxies  later assemble.

Detailed agreement between the standard model and the spectrum
of fluctuations in the microwave background gives us confidence that our basic picture of structure formation is correct. It predicts the first galaxy-size halos will accrete cooling baryons to form stars by a redshift of around 20 or so. Many expect intergalactic hydrogen was reionised by the first substantial generation of star-forming low mass galaxies. However, the detailed predictions are very uncertain. Indeed, intermediate mass black holes, visible as  miniquasars, are a theoretically compelling, although hitherto unobserved, component of the high $z$ universe that could also provide an important ionization source. Observations of the earliest galaxies, with the James Webb Space Telescope (JWST) and the new generation of 
extremely large telescopes (ELTs), and the nature of the intergalactic medium  at high redshift, as revealed by upcoming radio surveys, will be crucial in advancing our understanding of the astrophysical nature of reionisation.

Finally, although the interface between theory and observation is richest in the area of low redshift (z $<$ 2) galaxy formation, it is clear that the hierarchical dark matter driven assembly picture promoted in the 1980's and 1990's is too simplistic. Although new phenomenological ingredients are being added to the semi-analytical models in order to patch agreement between the theory and observations of new multi-wavelength 
surveys, it is clear we seek a break-through in our understanding.
All current galaxy formation theory is refined by "interplay" with the data: this is satisfactory for generating mock catalogues for future surveys but falls short  if one wants to arrive at a fundamental understanding of the physical processes governing
galaxy formation.

The missing ingredient seems to be an understanding of how star formation is regulated in galaxies. This is perhaps not really surprising. We cannot readily simulate how stars form in nearby clouds,  from Orion to Taurus, Inevitably this leads to new challenges when we try to cope with conditions that almost certainly were different at high redshift in the  early universe. 

A good example is the Kennicutt (1998) law, which relates the star formation rate to the gas content. Local data suggests this may be universal. However whether phenomenological modelling of the star formation rate, efficiency and initial mass function derived from studies of the nearby universe is relevant  to the high redshift Universe is far from certain. Some have argued that two modes of star formation are likely required: one to explain young, gas-rich, star-forming disks that are currently forming stars with low efficiency, and another to explain the presence of old, gas-poor massive spheroids that formed stars with high efficiency long ago.  

Another example is the initial stellar mass function (IMF). There are periodic attempts to argue that the IMF featured either more massive stars, or fewer low mass stars, at  high redshift or in extreme environments. The advantages of turning up the massive star frequency include accounting for the far infrared galaxy counts, rapid chemical evolution and boosting the number of ionizing photons for early  reionisation.  Systematically lowering the number of low mass stars may help with the colour-magnitude relation.

Finally, we cannot avoid remarking on the role of supermassive black holes in spheroid formation. There is little doubt that AGN are intimately connected to star formation. Unfortunately, neither theorists nor observers agree on the sign of the effect. Did black holes quench or trigger star formation? Most theorists favour quenching, but there are well-known counter examples. Did black hole growth precede or follow spheroid assembly? Most  but not all data suggests that black hole masses were supercritical (relative to the Magorrian relation) in the past. Making progress in this complex area at the interface of star and galaxy formation represents our third major challenge.

\section {Constituents of the Universe}

\subsection{Dark Matter}

Dark matter studies go back to the  pioneering efforts based on 
galaxy radial velocities by Zwicky for clusters and Rubin for galaxies.  This technique has to some extent been superceded by a more fundamental approach first predicted by Einstein, namely 
gravitational lensing by dark matter. Here we differentiate between `strong lensing', where the foreground lensing mass density is often sufficient to create multiple images of the background source,   and `weak lensing', the statistical distortion of the background population by large-scale structure. Although lensing is more often used to promote studies of  dark energy (see next section), it is important to realize its unique potential to track the properties of dark matter.

Weak gravitational lensing was only detected in 2000, but is now a 
highly-developed field. Dedicated  facilities are being constructed to exploit the effect,
and  significant progress is  expected from high quality imaging surveys. 
In the short term on the ground, these include PanSTARRS, the Dark Energy Survey,
the VLT Survey Telescope, VISTA and the Subaru HyperSuPrime Camera. Ultimately, in 
space, we may realize the NASA/DoE Joint Dark Energy Mission (JDEM) and the 
Dark UNiverse Explorer (DUNE) proposed for ESA's Cosmic Visions programme around 2015.

Existing facilities such as Subaru's SuPrimeCam imager, the CFHT Megacam
and the ACS onboard the Hubble Space Telescope have demonstrated the ability of weak lensing
to map the dark matter (DM) distribution, both in 2-D projection on the sky 
(Figure 1) and, via the use of photometric redshift data, reconstruction of the 3-D distribution. 
Comparing the distribution of DM, revealed by lensing, and the baryonic matter
as revealed by broad-band imaging, has verified the important assumption
that DM provides the basic `scaffolding' within which baryons
cool to form stars and assemble (Massey et al 2007a). 

\begin{figure}
\centerline{\hbox{
\epsfig{file=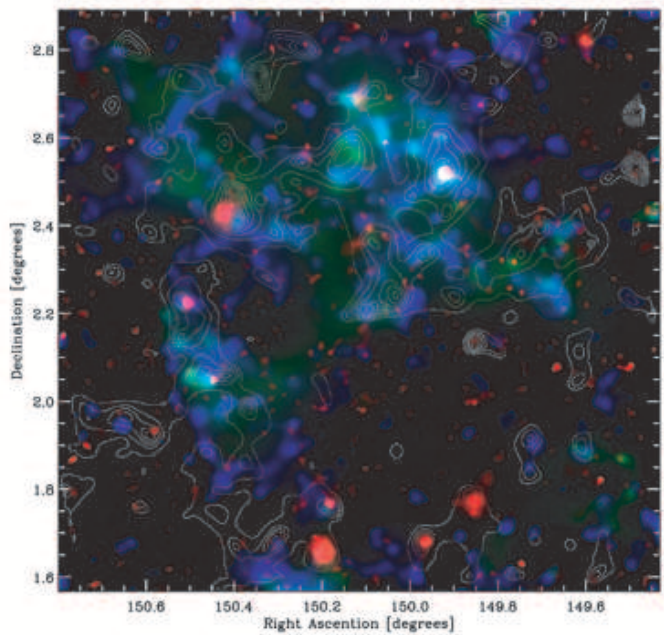, width=3.5truein}}}
\caption{\it The distribution of dark matter derived by weak lensing in the 
Hubble ACS COSMOS field (Massey at al 2007a). Such studies are 
influential in correlating the dark matter distribution (contours) with that of 
baryons, coded here according to optical emission (yellow), X-ray emission from
hot gas (red) and stellar mass (blue).}
\end{figure}

Strong lensing is likewise effective in providing important constraints on the distribution
of DM on small scales ($<$100 kpc). A key question is how can
extragalactic data provide constraints on the {\it nature} of the DM?
Numerical simulations based on the assumption that the DM is `cold'
and non-interacting suggest the radial profile of the DM should be
sharply peaked to small scales with a central density cusp $\rho\propto r^{-1}$ (Navarro, Frenk \& White 1998).
 
By combining measures of strong lensing in clusters with stellar dynamics
for the central cluster galaxy and its halo, Sand et al. (2004, 2007) have shown
how it is possible to isolate the distribution of DM on small scales for direct comparison
with the numerical simulations. Preliminary data, based on a few clusters,
suggests the DM is not as sharply peaked as predicted by CDM theory.
 Conceivably, this could arise
via prolonged gravitational interactions between the DM distribution and the
baryons, the latter for example receiving dynamical input by non-gravitational  forces.

Although few galaxy clusters have so far been studied with the detail required to 
make a convincing attack on this problem, the list of possible systems
to explore is quite large and, with sufficient effort, the outcome of
such a program could be very interesting.
As an extreme option, the presence of cores could imply the DM is not entirely cold.
For example, cluster dark matter cores might be the result of dwarf galaxy aggregation.
The apparent ubiquity and universality of cores in nearby dark matter-dominated 
dwarf galaxies has revived  interest in the possibility of warm dark matter consisting of 
sterile neutrinos in the few keV mass range.

The combination of strong and weak lensing data offers a powerful dynamical probe.
The SLACS project (Koopman et al 2006, Gavazzi et al 2007) has 
demonstrated that dynamical masses for elliptical galaxy lenses span a wide range in redshift. 
The sum of the  luminous and dark matter components `conspire' to form 
an isothermal distribution and the uniformity of the mass profile
with redshift is striking. Absence of any significant evolution in the inner 
density slopes suggest a collisional scenario in which gas and dark matter 
strongly couple during galaxy formation, leading to a total mass distribution 
that rapidly converges to dynamical isothermality (Figure 2).

\begin{figure}
\centerline{\hbox{
\epsfig{file=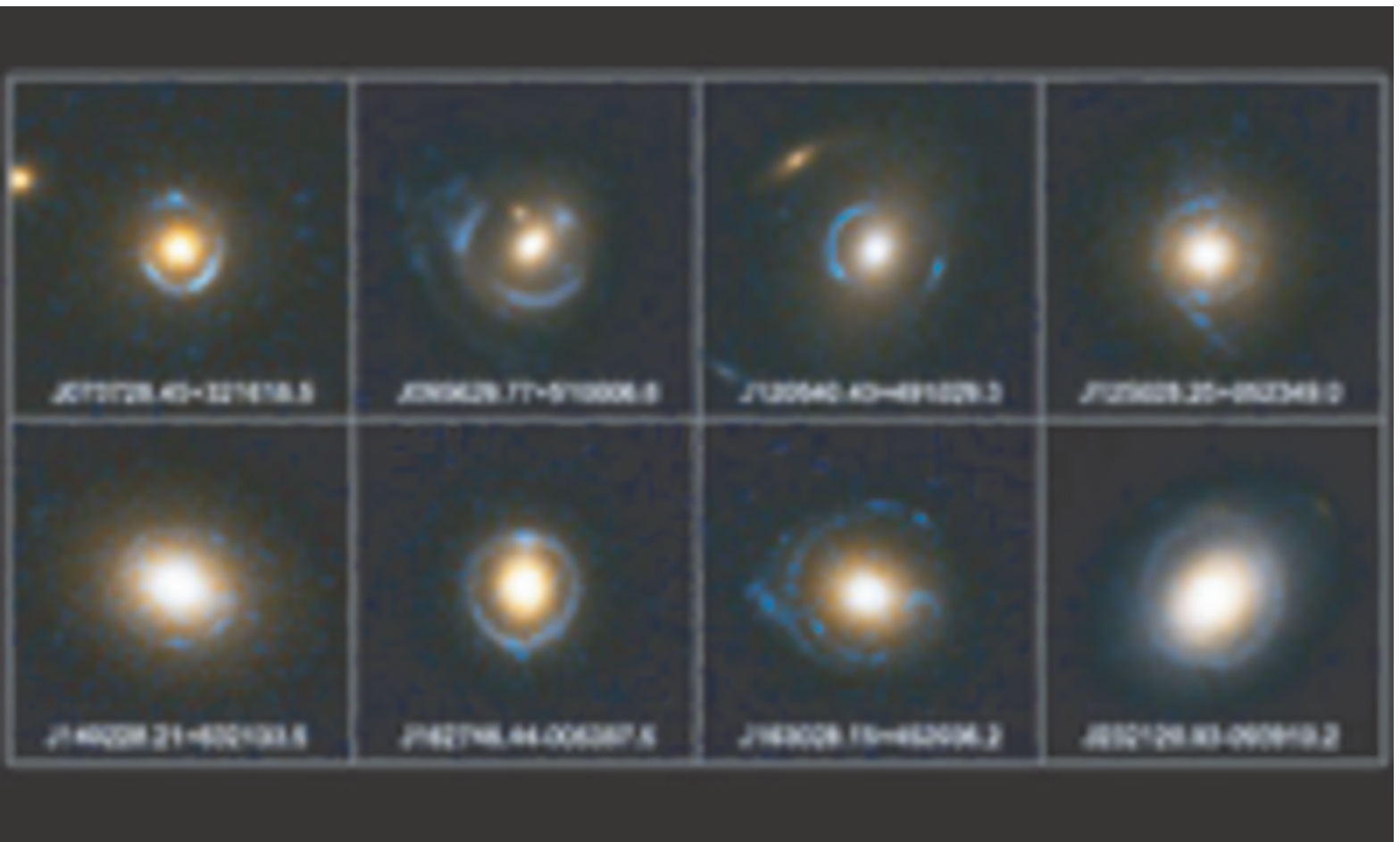, width=2.3truein}
\epsfig{file=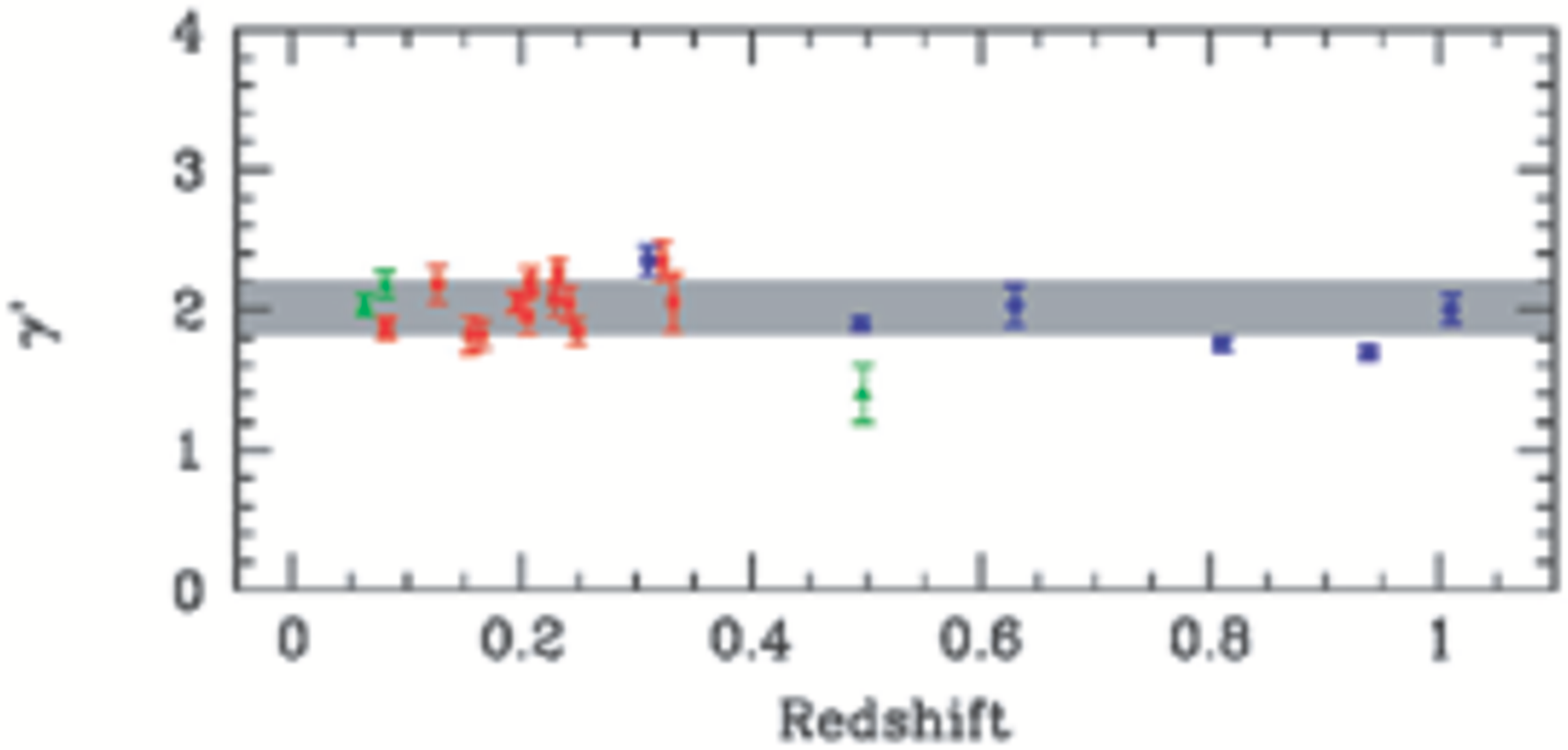,width=2.7truein}}}
\caption{\it (Left) The Sloan Lens ACS Survey led to a spectacular increase in the
number of known Einstein rings via ACS imaging of ellipticals whose spectra 
revealed additional line emission from background lensed sources. (Right)
The combination of strong lensing and stellar dynamics constrains the
inner slope of the logarithmic mass profile, $\gamma = d log\,\rho\,/\,dlog\,r$
which is remarkably close to isothermal at all redshifts (Gavazzi et al 2007).}
\end{figure}

\subsection {Dark Energy}

The past two years have seen unprecedented interest in developing new
facilities to address the perplexing question of the existence and nature of dark energy.
Dedicated ground-based telescopes and future space missions are being developed
or proposed with a variety of observational probes. But, given that there is no consensus 
on the physical basis of dark energy, one must exercise care in deciding
what is the optimal observational target and which
of the various observational probes is likely to be the most effective. More to the point, 
is a costly space observatory really required to make the necessary progress?

Recent studies in the US and Europe (Albrecht et al 2007, Peacock et al 2007)
have emphasized a step-wise approach where the most immediate target
is to determine, or otherwise, whether dark energy is consistent with 
the cosmological constant (a vacuum with an equation of state such that
$p= w\,\rho$ with $w$=-1). In the following, we explore the hurdles that
we must overcome with the four most
promising techniques in order to realize the necessary precision.

\subsubsection{Type Ia SNe}

This is the most well-developed method for probing dark energy. SNe Ia are
events rich in detail and the one parameter light curve "stretch" correction 
yields individual luminosity distances accurate to about 7\%.  The latest
results (Astier et al 2006), when combined with local baryonic oscillation
data for an assumed flat Universe, yield consistency with $w$=1 to
about 9\% accuracy.

Despite this progress, it is unclear whether SNe Ia are suitable for the required
next step in precision cosmology. Host-galaxy correlations (Sullivan et al 2006) demonstrate 
that SNIa explosions can occur with a wide range of delay times. This arouses the 
spectre of environmental and evolutionary biases. Could SNIa dimming with redshift,
generally accepted as the unique reliable measure of acceleration currently available,  be 
predominantly or even partly due to evolution? A significant UV dispersion 
has been seen in detailed spectra of $z\simeq$0.5 events (Ellis et al 2007)
which seems not to correlate with the light curves. Conceivably, this effect arises from
metallicity variations which may, ultimately, create a Òsystematic floorÓ  
particularly for the distant events that are essential for tracking the time dependence
of dark energy (Figure 3). Great efforts are needed to understand these possible
limitations, both via local high quality surveys and improved modeling of the
spectra.

\begin{figure}
\centerline{\epsfig{file=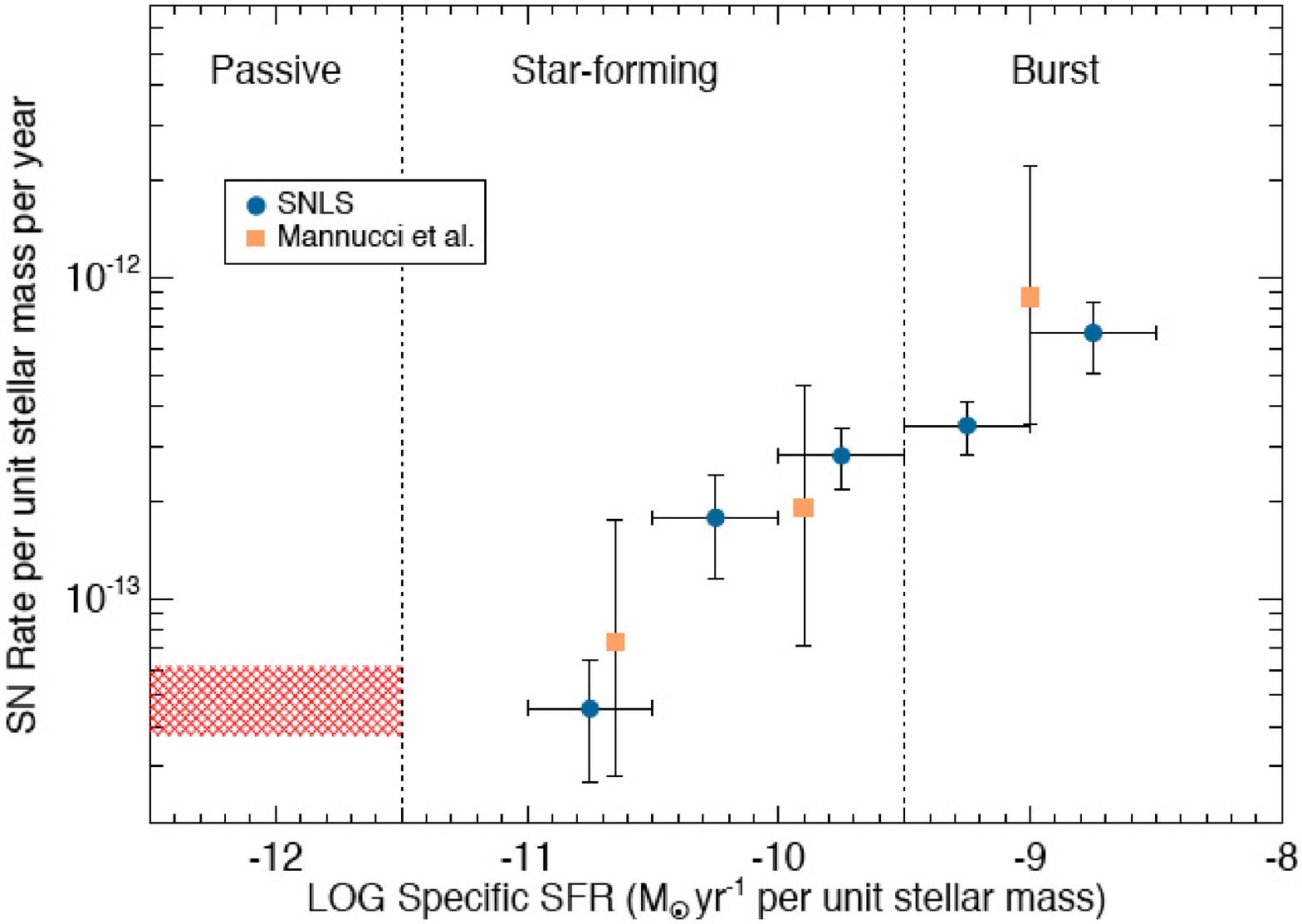, width=4.0truein}}
\centerline{\epsfig{file=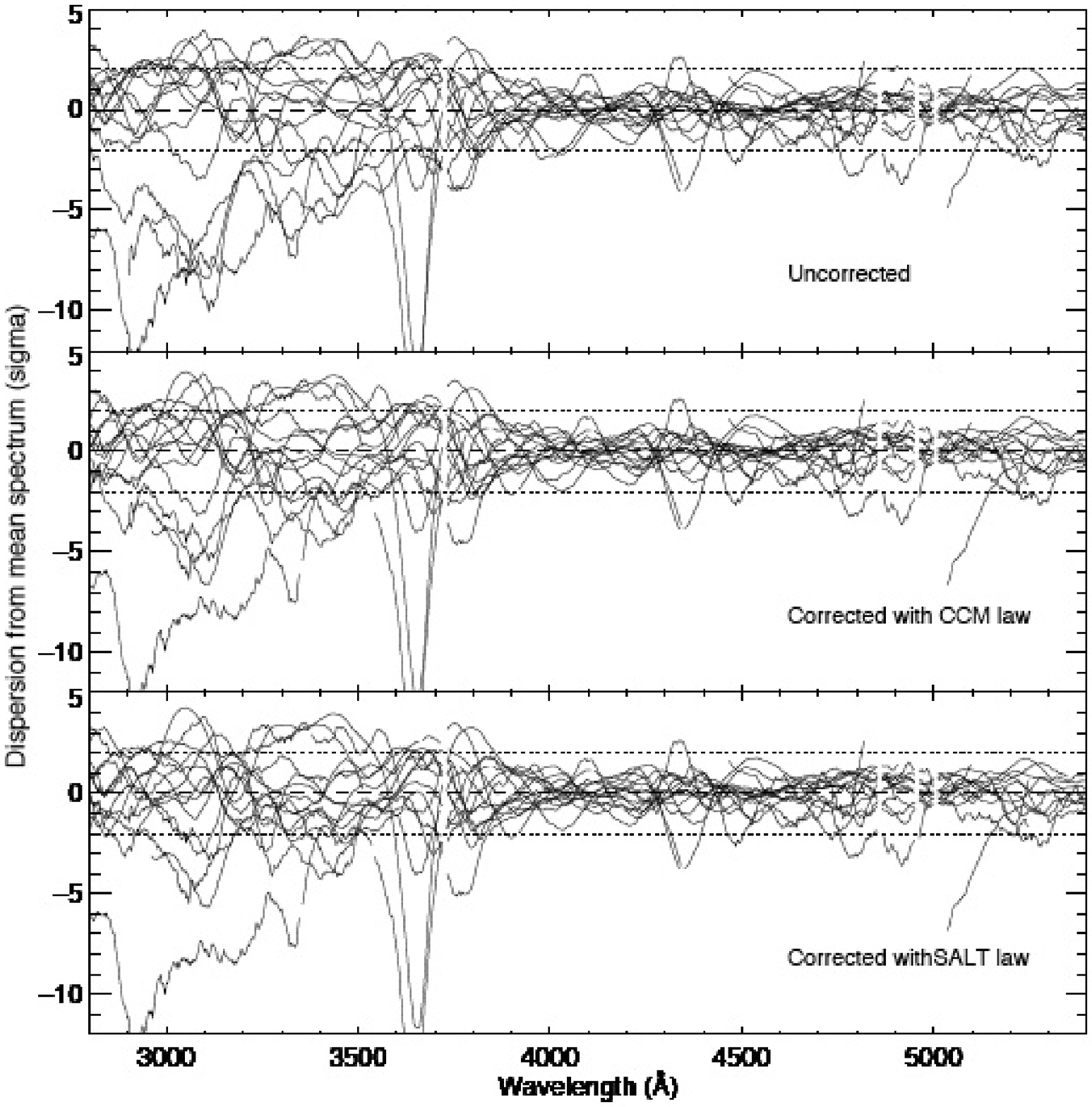, width=3.6truein}}
\caption{\it Challenges to the precision use of SNe Ia as distance indicators.
(Top) The Ia rate depends on the star formation history of the host galaxies
(Sullivan et al 2006). This suggests a range of delay times from birth to 
the explosion and possibly more than one progenitor mechanism. (Bottom) Spectral
dispersion at maximum light as a function of wavelength from the Keck survey
of Ellis et al (2007). The significantly increased UV dispersion is not removed
by standard calibration procedures and may indicate a new parameter is
at play (e.g. metallicity), which evolves with redshift.}
\end{figure}

\subsubsection{Weak lensing}
 
Weak lensing complements SNe as a measure of dark energy since, unlike
SNe which track the distance-redshift relation, lensing tracks the growth rate 
of density fluctuations. However lensing is less well-developed as a precision 
tool. Key issues include the accuracy with which the weak shear signal can
be calibrated and whether it is linearly recovered, and the reliability of
photometric redshifts, essential for `tomography', i.e. slicing the data in 
depth to recover the growth rate. The current state of the art, following
pioneering surveys at the Canada-France-Hawaii telescope, is consistent
with the SN results but with larger error bars (e.g. $w<0.8$ , Semboloni et al
2007).

Ground and  space-based observations each offer advantages.  Space 
offers increased depth, a higher surface density of useable background
galaxies and a more stable point spread function (Kasliwal et al 2007).  
Infrared coverage from space enables more precise photometric redshifts,
particularly for z$>$0.8 (Abdalla et al 2007).  Ground-based surveys offer
a more cost-effective areal coverage, the potential of associated wide field 
spectroscopy and associated non-cosmology science.

Although problems remain, the progress since weak lensing was detected
in 2000 is impressive. The Shear Testing Evaluation Program (STEP),
a collaborative effort using simulations of known shear, is being used to test 
the various image processing algorithms (Heymans et al 2006, 
Massey et al 2007b).  Although most of the algorithms in regular use do not 
yet recover shear at the necessary precision, one can hope to learn
from these and related developments.
 		 
 \subsubsection{Baryon Acoustic Oscillations}
 
 Baryon ÒwigglesÓ are an imprint on the dark matter of the gravitational potential 
 variations due to acoustic oscillations of baryons  from the epoch of 
 matter-radiation decoupling. This  relic of the horizon scale in the galaxy 
 power spectrum may be the perfect `cosmic rulerÕ. As the signal is weak
 and on very large scales ($\simgt$ 120 Mpc), well-sampled redshift surveys
 in huge volumes are needed to achieve the necessary precision.
 A major hurdle to be overcome will be the biasing of galaxies relative to 
 dark matter. This can, in principle, be modeled via numerical simulations.
 
 The most substantial project currently underway is WiggleZ at the AAT 
 (Glazebrook et al 2007) which plans to obtain more than 200,000 emission
 line redshifts  to $z\sim 1$, thereby constraining $w$ to 10\% (Figure 4). A similar
 SDSS-based extension is likely to exceed this precision by a factor of 2
 or so. Such accuracy ($\sim 5\%$) cannot easily be superceded with imaging 
 surveys that utilize photometric redshifts, which have intrinsic limitations 
 especially at $z\simgt 1.$  Thus the next generation instruments
 for BAO studies will necessarily have to be quite ambitious. Gemini/Subaru are 
 funding  conceptual designs for a 1.5 square degree field multi-fiber 
 spectrograph (WFMOS). Although motivated in part as a dark energy
 experiment, such an instrument would represent a logical 8m successor
 to SDSS and 2dF and offer much valuable science, particularly in
 studies of Galactic archeology. 
 
 A longer-term development involves a space-based spectroscopic survey such as DESTINY (a grism-type instrument proposed for NASA JDEM) or SPACE (a microlenslet instrument proposed for ESA Visions), which will obtain, in the latter case, up to a billion galaxy spectra. Whether ground or space-based, one is looking at huge investments,
$\sim \$50 M$ or $\sim \$500 M $ respectively, to obtain a factor of order 2 improvement in precision for $w$. The possible pay-off. which cannot be overstated,  is a  signature of new physics such as $w\neq -1$. The downside is that one may just end up confirming Einstein's cosmological constant.
 
\begin{figure}
\centerline{\hbox{
\epsfig{file=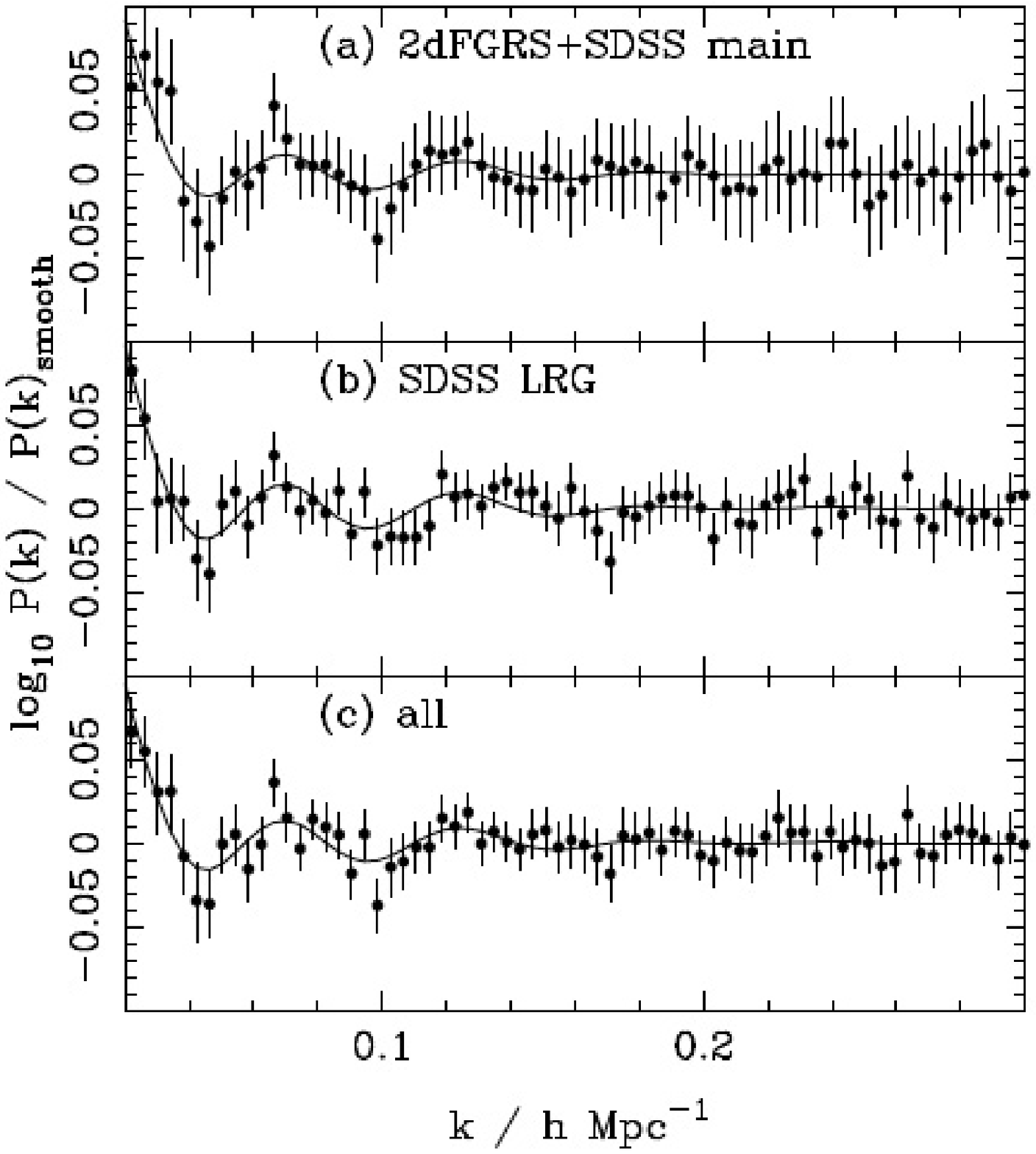, width=2.5truein}
\epsfig{file=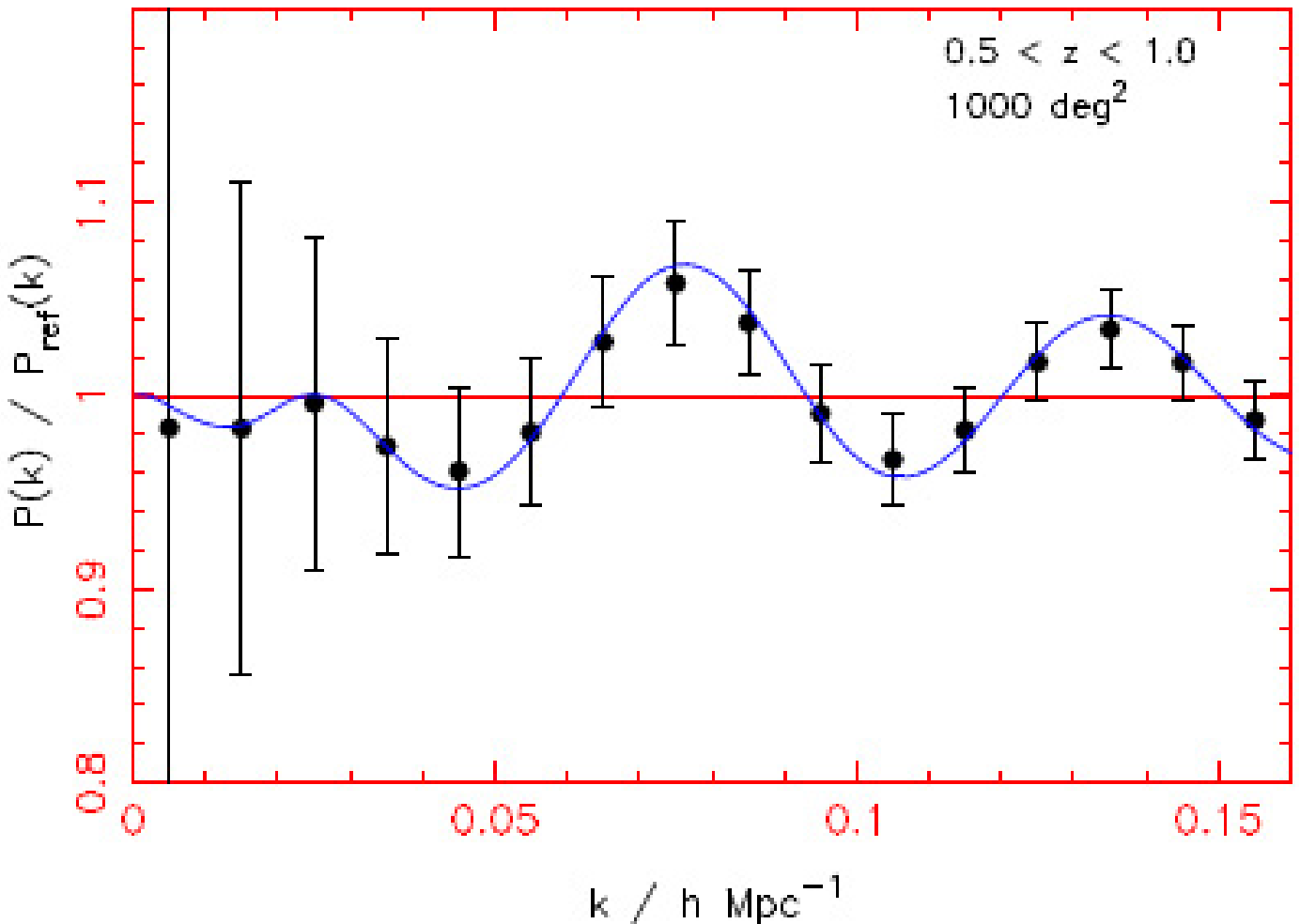, width=2.6truein}}} 
\caption{\it How baryonic oscillations work: (Left) Detection of baryonic
oscillations in the local 2dF and SDSS redshift surveys (Percival et al
2007). The galaxy power spectrum is divided by a smooth function to reveal 
oscillations on large scales which represent the expanded relic of
the horizon scale at recombination. (Right) With sufficient data, the 
wavelength of the oscillations acts as a standard ruler enabling the 
expansion history of the Universe to be directly detected (simulation
of AAT Wigglez project, Glazebrook et al 2007).}
\end{figure}
 
\subsubsection{Galaxy clusters}

Large samples of clusters are planned from X-ray and Sunyaev-Zeldovich 
(SZ) surveys. The path to dark energy again arises from the growth
of structure. The most direct route is via the cluster number dependence on mass
and redshift, $N(M,z)$, which constrains $w$ via the perturbation growth 
factor.  A key requirement is thus the mass of each cluster.  Calibrating
masses to the necessary precision, and with adequate survey uniformity, 
via the X-ray luminosity or temperature or the SZ decrement is the challenge. 
Ideally, one combines both approaches once the systematics are understood.

Current planning includes a survey of $10^5$ X-ray clusters on eROSITA 
(projected launch 2010). The claimed precision on $w$ is  5\% (Haiman et al. 2005), 
assuming a mass calibration that is precise to 1\%. Much progress on calibration
accuracy will be necessary to realize this goal given tthat he current mass uncertainties
are of order 10\% for simulated data (Nagai, Kravtsov and  Vikhlinin 2007).

\section{First Light and Cosmic reionisation}

We now turn to our second frontier topic: the location of the first star-forming 
galaxies and evaluation of their role in bringing the so-called `Dark Ages' to an 
end and reionizing the intergalactic medium. We have very few indicators
of when cosmic reionisation occurred but evidence is accumulating that
star-forming galaxies were the responsible agent. 

\begin{enumerate}

\item{} The polarization - temperature cross-correlation derived from the WMAP
Three Year data set suggests a redshift window of 7$<z<$20 for the location of the 
bulk of the scattering electrons (Spergel et al 2007).

\item{} The assembled stellar mass at $z\simeq$5-6 (Eyles et al 2006, Stark et al 
2007a) indicates a substantial amount of star formation at earlier epochs, perhaps
sufficient to explain reionisation. Many of the more massive galaxies which can
be studied in detail at this epoch show `Balmer breaks' - signatures of old ($>$100
Myr) stars in galaxies (Eyles et al 2005, Figure 5).

\item{} Carbon, produced only in stellar cores, is present in absorption in the
highest redshift QSO spectra and seems ubiquitous in the intergalactic medium
out to $z\simeq$6 (Ryan-Weber et al 2007). This suggests an early period of 
enrichment from supernovae.

\end{enumerate}

The negatives are that one has no idea of the escape fraction for ionizing photons in the 
first galaxies and that  there is at least one plausible  alternative source of ionizing photons. This consists of intermediate mass black holes (IMBHs), which act as miniquasars  and are prolific sources of ionizing photons at very early epochs.
They must be present in considerable numbers  in the early universe if one is to understand how supermassive black holes were in place by $z\sim 6$ 
as evidenced by the presence of ultraluminous quasars. Theoretical arguments suggest that the first generation of dissipating gas clouds at $z\sim 10$ could as easily form IMBHs as population III stars, and indeed probably form both. Confirmation of such a high redshift population of non-thermal ionising sources could eventually come from a combination of x-ray background, 
high-$\ell$ CMB and LOFAR observations.

\begin{figure}
\centerline{\hbox{
\epsfig{file=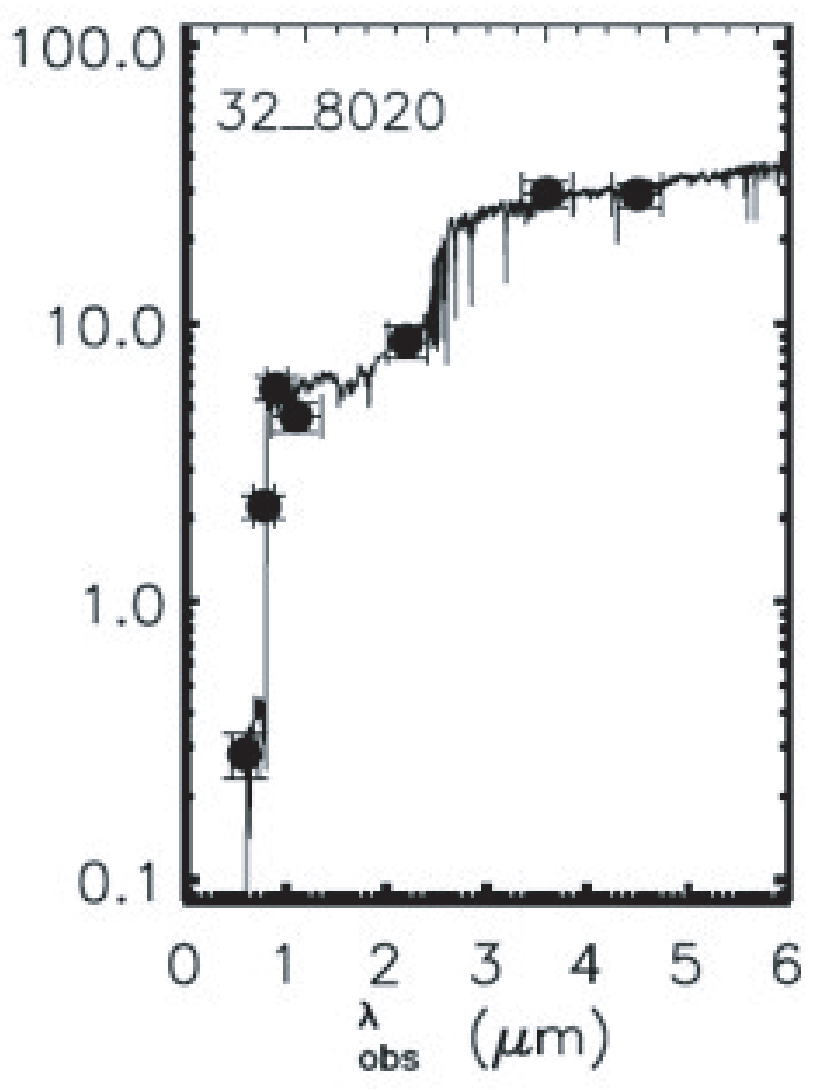, width=2.2truein}
\epsfig{file=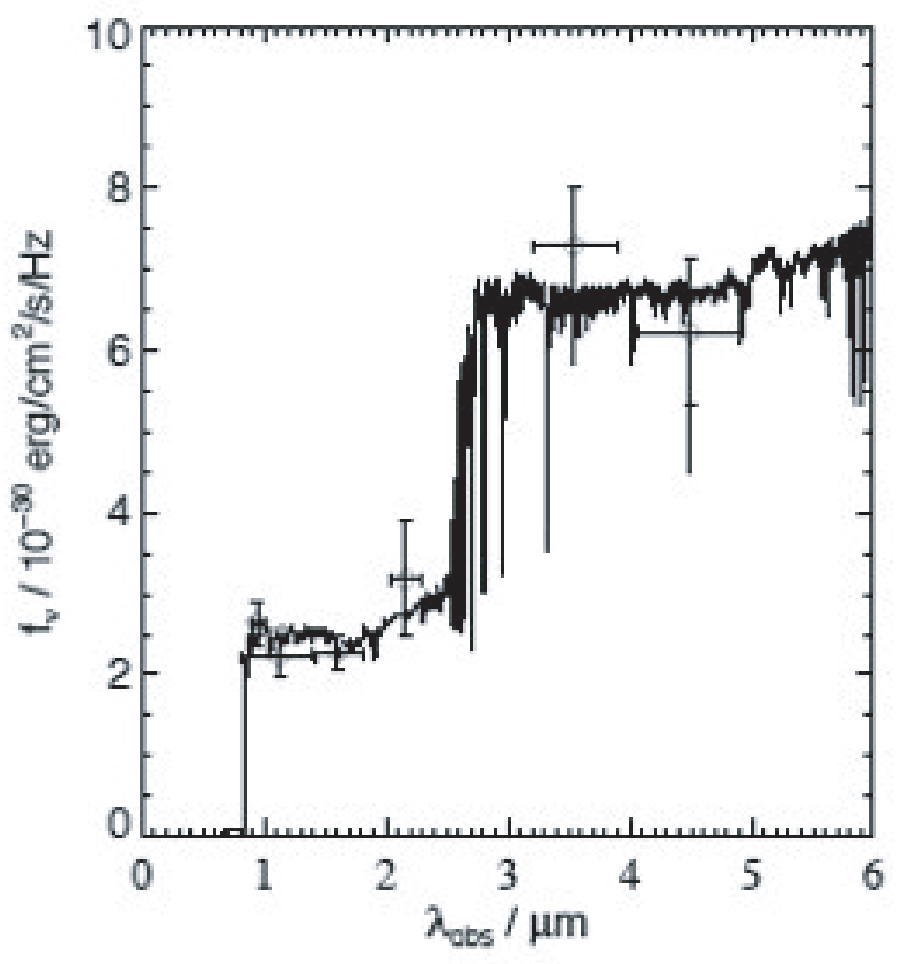, width=2.6truein}}} 
\caption{\it Spectral energy distribution of two spectroscopically-confirmed 
high redshift galaxies with established ($>$100 Myr) stellar populations as
revealed via their significant `Balmer break' at 2-3 $\mu$m with Spitzer.  
(Left) A $z$=5.55 galaxy with a stellar mass of 1.1 10$^{11} M_{\odot}$ from
the survey of Stark et al (2007a). (Right) A $z$=5.83 galaxy with a stellar mass
of 2-4 10$^{10} M_{\odot}$ from the survey of Eyles et al (2006).}
\end{figure}

Given the evidence for early activity, several groups are now probing the
era  $z>$7 for the first glimpse of the sources responsible for reionisation
(for a review, see Ellis 2007). Unfortunately, prior to the launch of JWST and the commissioning
of the next generation of 30-40 meter telescopes, progress will inevitably  be very
slow. A key question is whether the bulk of the sources responsible for
reionisation are luminous and rare, or abundant and feeble. As there
is growing evidence over 3$<z<$7 that the luminosity function of both
continuum `dropouts' and Lyman $\alpha$ emitters is steepening at the
faint end with redshift (Bouwens et al 2007, Kashikawa et al 2007), it
seems prudent to search for an abundant population of sources whose 
star formation rates are less than 1 $M_{\odot}$ yr$^{-1}$.

One possible route to discovering such faint sources, if they are very 
numerous, is via strong gravitational lensing. A rich cluster of galaxies can
magnify selectively distant sources by factors of $\times$25 along their 
`critical lines'. As several distant sources have been successfully recovered
using this technique (Ellis et al. 2001, Kneib et al. 2004), a number of
groups have begun searching for faint $z>$7 galaxies in these regions
(Figure 6). Via this approach, Stark et al. (2007b) have published 6 candidate 
Lyman $\alpha$ emitters, some of which they claim lie in the redshift 
range 8.6$<z<$10.2 with (unlensed) star formation rates as low as
0.2 $M_{\odot}$ yr$^{-1}$. Since the lensed searches cover only a
small area (0.3 arcmin$^2$ in Stark et al.'s case), if even a fraction of 
these sources are truly at these redshifts, the abundance of low
luminosity galaxies is within striking distance of that required for cosmic
reionisation.

\begin{figure}
\centerline{\hbox{
\epsfig{file=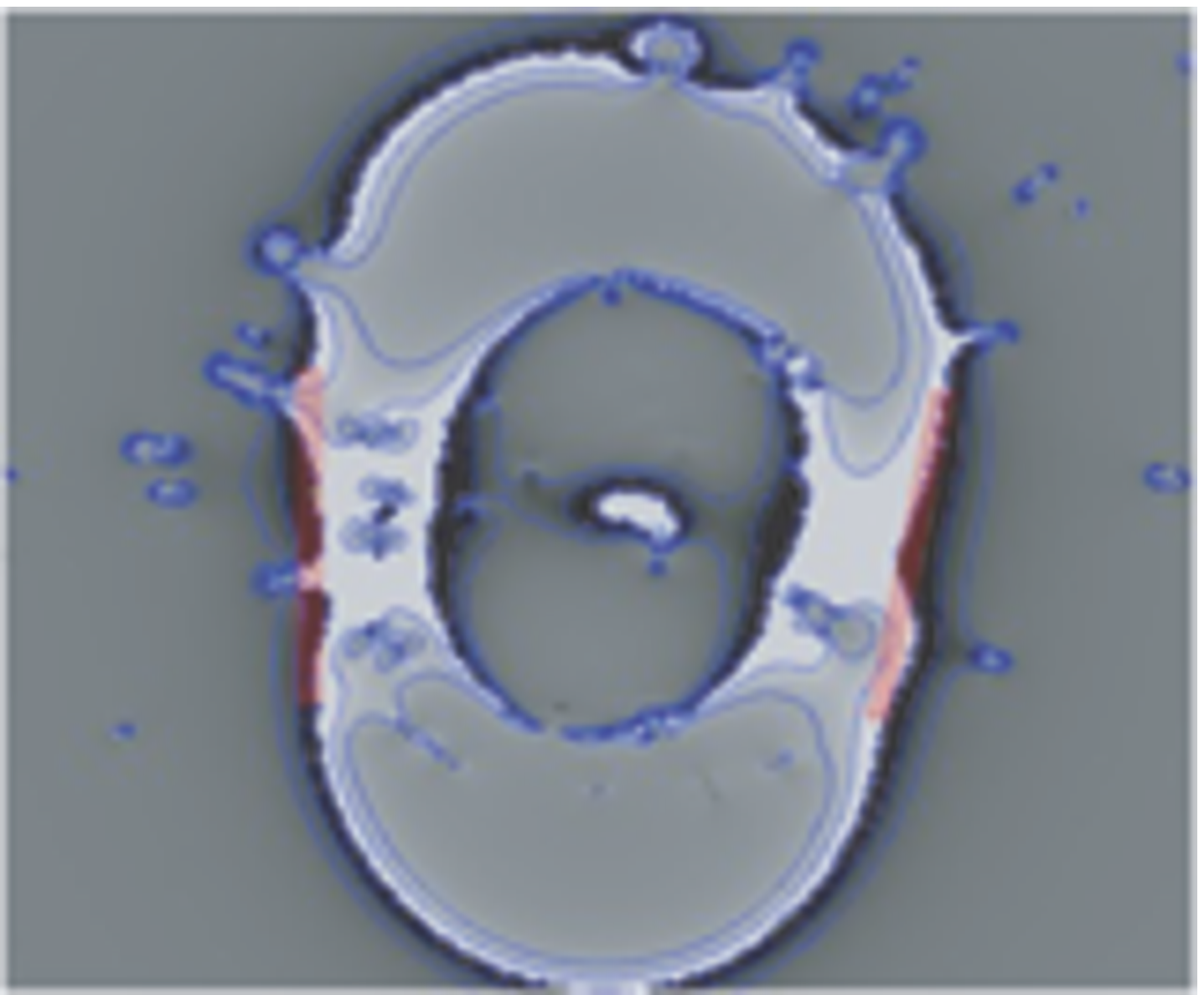, width=2.5truein}
\epsfig{file=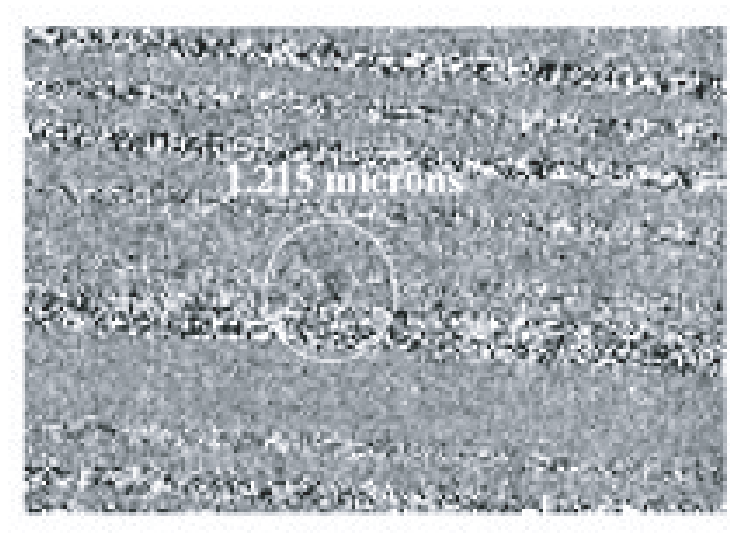, width=2.6truein}}}
\caption{\it Probing to high redshift with gravitational lensing.
(Left) Magnification map for the rich lensing cluster Abell 370. Two 
curves of maximum magnification (`critical lines') encircle the 
cluster. By searching in a blind manner spectroscopically within 
these narrow regions (red lines), highly magnified line emitters from $z>$7 
can be located. (Right) 2-D spectral image of one of 6 candidate 
Lyman $\alpha$ emitters at $z$=10.0 found in the Keck infrared 
survey of Stark et al (2007b).} 
\end{figure}

\section {Galaxy formation}

Our final challenge is the search for a full understanding of how
the star-forming galaxies that we see at $z\simeq$2-3 assemble
into the varied forms which we see locally.

\subsection{Disk galaxies}

The disk mode of galaxy formation is motivated by the gravitational instability 
of gas-rich disks. Disk galaxies form late, slowly and inefficiently. The inefficiency 
is plausibly interpreted in terms of momentum input into the interstellar medium  
from expanding supernova shells. Supernovae provide the quantitative amount of 
feedback required to account for the star formation rate observed in  disk galaxies.
Star formation is sustained by a combination of disk self-gravity and  supernova feedback.
The self-gravity drives large-scale instability provided the disk is cold. 
Continuing accretion of cold gas guarantees the gaseous disk continues to 
fragment into large molecular cloud complexes. These accrete smaller clouds, 
cool and themselves fragment into stars.  Calculation of the star formation rate 
is complicated by the role of magnetic fields, whose pressure also  plays a role 
in supporting clouds against collapse and decelerating star formation. The 
magnetic fields themselves are ionisation-coupled to the cold gas, and the ionisation 
fraction is itself due to  radiation from massive stars. The situation is further 
complicated by the fact that many clouds are supported by turbulent pressure, 
and the turbulence driver is not well understood. 

Despite these non-linear complications, the global star formation rate in galactic disks
is described by a remarkably simple formula:
$$SFR=0.02 \, \rho_{gas}\, / \,t_{dyn}.$$
This simple model fits a wide range of data, including quiescent and star-bursting 
galaxies, and even the individual star-forming complexes in M51 (Kennicutt et al. 2007) as well as 
submillimeter galaxies and ULIRGs at $z\sim 2$ (Bouche et al. 2007). Evidence for local cold gas feeding 
is found in nearby disk galaxies that have extensive HI envelopes.

While one can phenomenologically understand star formation in disks, there is no 
successful model for disk formation. All attempts to date lead to overly massive or concentrated 
bulges. Detailed studies of self-gravitating hydrodynamic collapse with realistic initial conditions 
find that while the initial specific angular momentum matches that of observed disks, 
some 90\% is lost to the halo during collapse. Late infall implies gas-rich halos, which in turn require large SN-heating efficiencies (Governato et al. 2007)
and result in predicted large x-ray  luminosities (Sommer-Larsen 2006). The implied large amounts of hot gas are at best only rarely seen for massive disk galaxies, where coronal absorption lines set strong constraints (Wang et al. 2007).

In fact, Fraternali and Binney (2007) argue that the HI 'beards"  found in nearby spiral galaxies such as NGC 891 and NGC 2403 are analogous to galactic high velocity clouds and are signposts of a substantial and otherwise mostly hidden 
halo gas accretion rate that interacts with supernova-driven galactic fountains. While the required  accretion rate is consistent with LCDM expectations and is 
comparable to the star formation rate, the infalling gas is required to have low angular momentum in contrast with expectations from simulations (Gottloeber and Ypes 2007).

\subsection {Early-type galaxies}

There is a long-standing debate about how early-type galaxies formed. 
Was the process primarily monolithic or hierarchical? The response seems to 
center on understanding star formation as opposed more simply to mass assembly.

The following observations argue for a monolithic origin.
\begin{itemize}

\item{} $[\alpha/Fe]$: The ratio of SNII-generated alpha elements to iron, mostly 
produced by SNIa, is a powerful clock that demonstrates star formation in massive 
galaxies proceeded more rapidly than in  lower mass systems (Thomas et al. 05).

\item{} {\it Star formation time:} SED analysis of distant galaxies favours 
systematically lower specific star formation rates for increasing stellar mass. 
This supports downsizing: massive galaxy formation preceded low mass 
galaxy formation. Star formation in massive galaxies was complete before the peak in cosmic star formation history at $z\sim 1.5$ (Papovich et al. 2006).

\item{} {\it Fundamental Plane:} The low dispersion in the fundamental plane for 
elliptical galaxies indicates a narrow spread in ages at a given stellar mass.

\item{} {\it Color-magnitude relation:} The low dispersion argues for a 
conspiracy between age and metallicity. 
\end{itemize}

A hierarchical origin is supported by the following arguments.
\begin{itemize}
\item{} {\it  Environment:} The dependence of morphological type on environment,
especially mean density and cluster membership, argues for formation via a merging history.

\item{} {\it Major mergers:} Ultra-luminous infrared and submillimetre galaxies 
invariably show evidence of triggering by major mergers.

\item{} {\it Morphological evolution:} Deep surveys show that the Hubble sequence is 
essentially unchanged to $z\sim 1$, but rapid evolution in both size and 
morphology occurs at higher redshift.

\item{} {\it Star formation history: } The color-magnitude relation conspiracy may be 
broken with GALEX data, which however leads to new issues concerning the gas 
supply needed to account for the residual star formation found in 30\% of nearby 
early-type galaxies. Significant minor mergers provide a plausible explanation.
\end{itemize}
\begin{figure}
\centerline{\hbox{
\epsfig{file=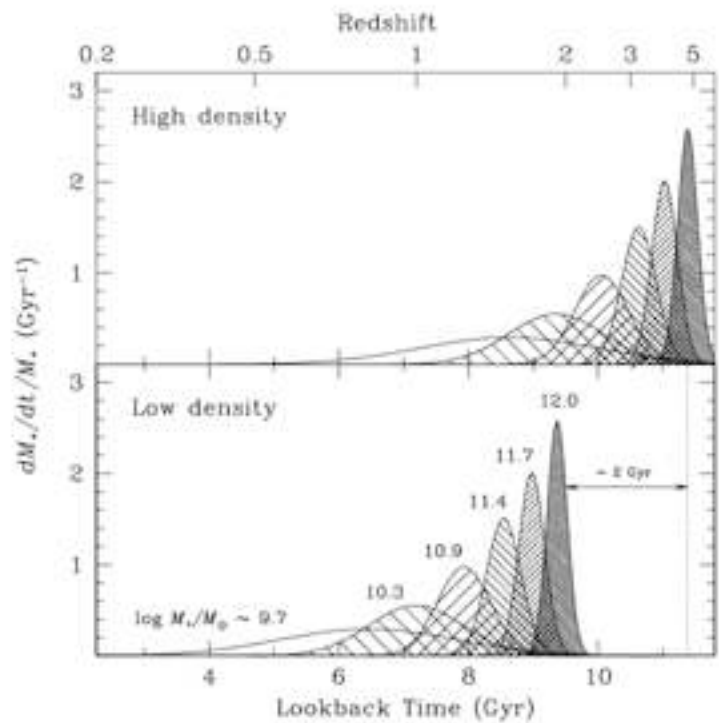,width=4truein}}}
\caption{\it The evidence from the abundance ratios of alpha elements to iron 
 demonstrates downsizing in star formation history (Thomas et al. 2005).  Massive early-type galaxies 
 systematically form earlier (from SED fitting) and have shorter star formation time-scales (from
 $[\alpha/\rm Fe]$).
 }
\end{figure}

Reality most likely is a combination of both hierarchical and monolithic scenarios. The 
former controls disk formation, the latter seems appropriate to massive spheroid 
formation. However even monolithic star formation is reconcilable with hierarchical 
assembly of dark matter and gas, although stars could only track hierarchical 
aggregation to a much lesser extent. Most stars in massive early-type galaxies 
must be formed over a period of order a dynamical time, and hence monolithically.
However a pure monolithic scenario is incompatible with the observed cosmic star formation 
and galaxy assembly history at high redshift. It also results in far too low a rate of ionizing photon production at $z\simgt 6.$
 
 \subsection{Feedback}
 
Semi-analytical modelling gives an excess both of small and massive galaxies relative 
to the observed luminosity function. Resolving the galaxy luminosity function problems 
is possible with astrophysical feedback. There is an active debate as to the optimal 
feedback mechanism. A majority view is that AGN-driven outflows provide substantial
feedback in the early phases of massive spheroid formation 
(Bower et al. 2006, Croton et al. 2006]).
This provides a means of quenching star formation by heating the gas reservoir and
dispersing it. AGN outflows provide a heating mode that is effective on the intracluster gas at low redshift. It is essential to avoid continued star formation via the accreting gas. Indeed, there is evidence for entropy injection, both from the observed entropy profiles, the cooling cores that cool at unexpectedly low rates, and the presence of AGN-driven bubbles.

A clue that favours blowout comes from the correlation between central black hole mass
and spheroid velocity dispersion. If gas accretion fuels both black hole growth and star
formation, exhaustion of the gas reservoir occurs when the black hole mass is high
enough for blowout to drive out the residual gas. Collimation could affect this argument,
although jet instability in an inhomogeneous protogalaxy would rapidly degenerate into 
a cocoon. Blowout occurs, star formation terminates and  the SMBH-$\sigma$  
relation saturates. The resulting momentum balance condition between self-gravity of 
the protogalactic gas and Eddington-limited outflow yields a relation between
supermassive black hole mass and spheroid velocity dispersion that fits the
observed normalisation and slope of the correlation (Silk and Rees 1998).
Correlations
between ultraluminous starbursts, winds and AGN  may be a manifestation of this 
process at high redshift.  Observations of x-ray selected quasars show that AGN feeding and SMBH growth  coevolve and peak along with the cosmic star formation history at $z\sim 2$ (Silverman et al. 2007). 
 .
\subsection{Downsizing}

A major enigma has come from the phenomenon of galaxy downsizing. In one study, 
the Palomar-DEEP2 galaxy stellar mass functions over $0.4<z<1.4$ have been 
obtained  using rest-frame U-B colors as a discriminant (Bundy et al 2006). A threshold galactic stellar 
mass is apparent above which there is no star formation. This mass threshold 
increases from $10^{11}\rm  M_\odot$ at $z\sim 0.3$ to $10^{12}\rm  M_\odot$ 
at $z\sim 1$ (Figure 8).  The most massive galaxies are already in place by $z\sim 3.$
The evidence from NIR stellar mass  estimates
 demonstrates that downsizing in mass assembly over a broad range in redshift (Perez-Gonzalez et al. 2007).
Parallel studies focussing on X-ray selected AGN similarly 
find that AGN luminosities have a luminosity threshold that increases to beyond 
$z\sim 2$ from the weakest to the strongest AGN (Hasinger, Miyaji and Schmidt  2005).
 
\begin{figure}
\centerline{\hbox{
\epsfig{file=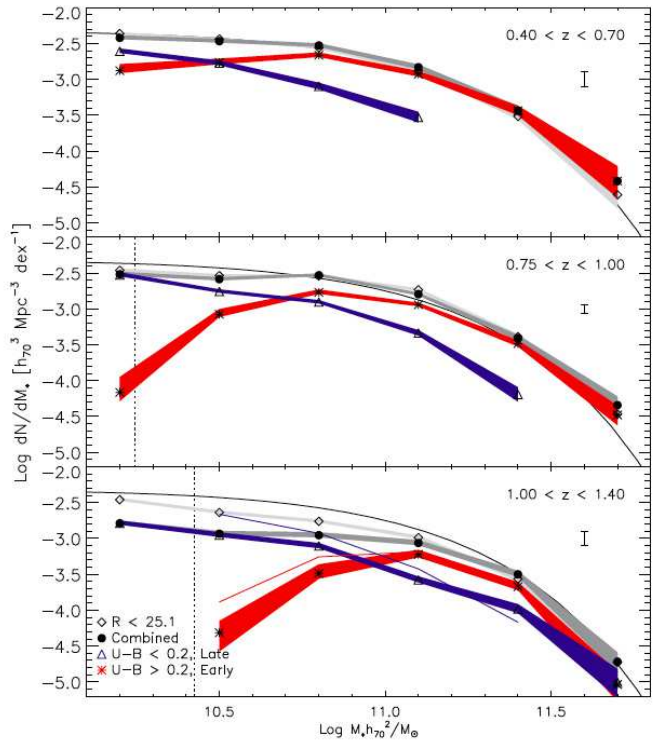,width=3.5truein}}}
\caption{\it A detailed analysis of downsizing in the Palomar-DEEP2
survey (Bundy et al 2006). The panels show the stellar mass function
in 3 redshift intervals partitioned according to a measure of star formation:
red - quiescent, blue - active. An evolving mass threshold is apparent above
which star formation is quenched.}
\end{figure} 

The conclusion is that both early-type galaxies and SMBH masses are 
anti-hierarchical, that is to say, the most massive objects are present early, whereas 
the least massive are absent at early epochs. This phenomenon was {\it not}
predicted by the pioneering models on galaxy formation in a CDM-dominated universe.
Since low mass halos predominated 
early, star formation must have been more efficient in the most massive galaxies.
Why is there downsizing of both massive spheroids and AGN, with a similar rise 
between redshift 0 and $\sim 2$? 

There are alternative  schools of thought on the resolution of this paradox.
Gravitational heating is environmentally induced and mostly affects massive galaxies which are found in the most massive halos. This can terminate late star formation, while the environment provides intense gas feeding and fueling provided by minor mergers (Khochfar and Ostriker 2007). This scenario gives downsizing, at the price of prolonged star formation.  Cooling primarily affects the lower mass halos where there is late infall of cold gas along filaments and star formation is correspondingly more efficient (Dekel \& Birnboim 2007). This helps explain the scaling relations (colour, magnitude, metallicity), where the relations flatten above a characteristic galactic stellar mass of $\sim 3\times 10^{10}\rm M_\odot$
(Tremonti et al. 2004).

The major, and prevalent, rival view appeals to AGN heating of the ICM to suppress cooling flows at late epochs (Bower et al. 2006; Croton et al. 2006). 
Early AGN activity is Eddington-limited, and the cooling times are shorter. Massive galaxies form early as a consequence. AGN feedback at late times primarily suppresses gas cooling and star formation in  massive halos. This  terminates both massive black hole growth and spheroid star formation, and 
results in downsizing as the late cold gas supply is quenched.  One obtains simultaneous  downsizing both for AGN and for the massive host galaxies. However the star formation time-scale in massive galaxies again is long. Neither model gives the short time-scales seen in SED modelling of massive galaxies and especially in the alpha element ratios. Nor do the models account for  the low dispersion in scaling relations such as the fundamental plane.
\begin{figure}
\centerline{\hbox{
\epsfig{file=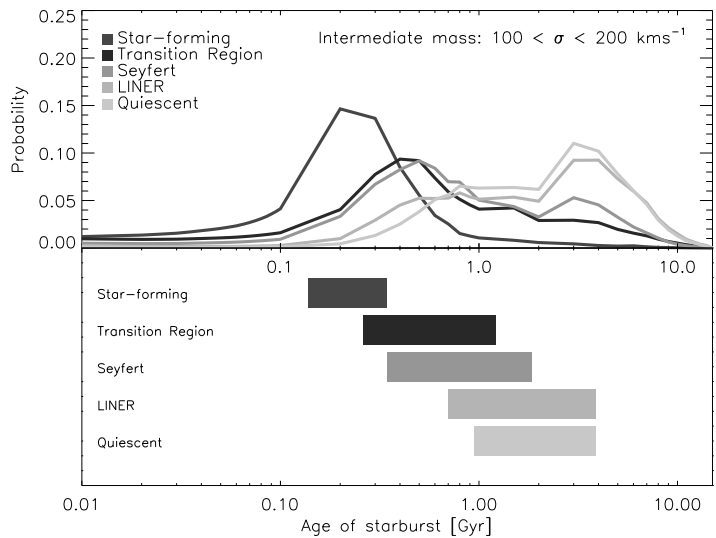, angle=90, width=3.5truein}}}
\vskip 0.2cm
\caption{\it The link between starbursts and AGN is demonstrated by this time sequence for low and intermediate masss galaxies, with time delay indicated from onset of star formation for different types of activity in early-type galaxies (Schawinski et al. 2007)}
 \end{figure}

What could be the missing link? One hope is that better simulations at higher resolution may 
help resolve this discrepancy.  But additional physics most likely is needed.
AGN triggering could  provide a solution.
If indeed the Eddington luminosities are high enough to drive the gas out, the 
AGN-driven outflows  could interact with the gas strongly enough to trigger star 
formation  by jet-driven cocoon overpressuring of  massive interstellar clouds. 
The jet drives a series of weakly supersonic shocks into the cocoon, which fills with disrupted cloud gas. The cores of the massive clouds survive and are overpressured. Moreover, the turbulent backflow overpressures the protospheroid and protodisk. The enhanced pressure induces gravitational instability and star formation (Antonuccio \& Silk 2007).  The jet/cocoon time-scales associated with compact radio sources are short, of order $10^6$ yr. Moreover the interstellar gas kinematics in high redshift radio galaxies are seen in some cases to be controlled by cocoon-driven turbulence (Nesvada et al. 2007).
The time-scale of triggering, associated with the jet/cocoon propagation speed  and necessarily short compared to the dynamical time-scale, results in an 
enhanced rate of star formation. Downsizing is a natural corollary if the 
the  SMBH feeding, jet power and outflow strength are a strong function of 
galaxy  and SMBH mass, as inferred observationally. For example, 
 the radio power is found to correlate with the Bondi accretion rate for an 
 x-ray sample of massive radio galaxies (Allen et al. 2006).
 The next phase would be quenching, aided and abetted by the boosted star formation and SN rate. It is only at this later stage that the AGN may be observable as an x-ray source. In the earliest stage  there is  correlated AGN and star formation activity. This phenomenon may be visible in the mid-IR for massive galaxy precursors such as  SMGs and ULIRGs at $z\sim 2$  (cf. Pope et al. 2007; Sajina et al. 2007). This is a chicken and egg problem of course: it remains to be shown which is the driver. Possibly it may be neither: theory suggests that a major merger may feed both the AGN and drive the starburst, as discussed below.
 
 \begin{figure}
\centerline{\hbox{
\epsfig{file=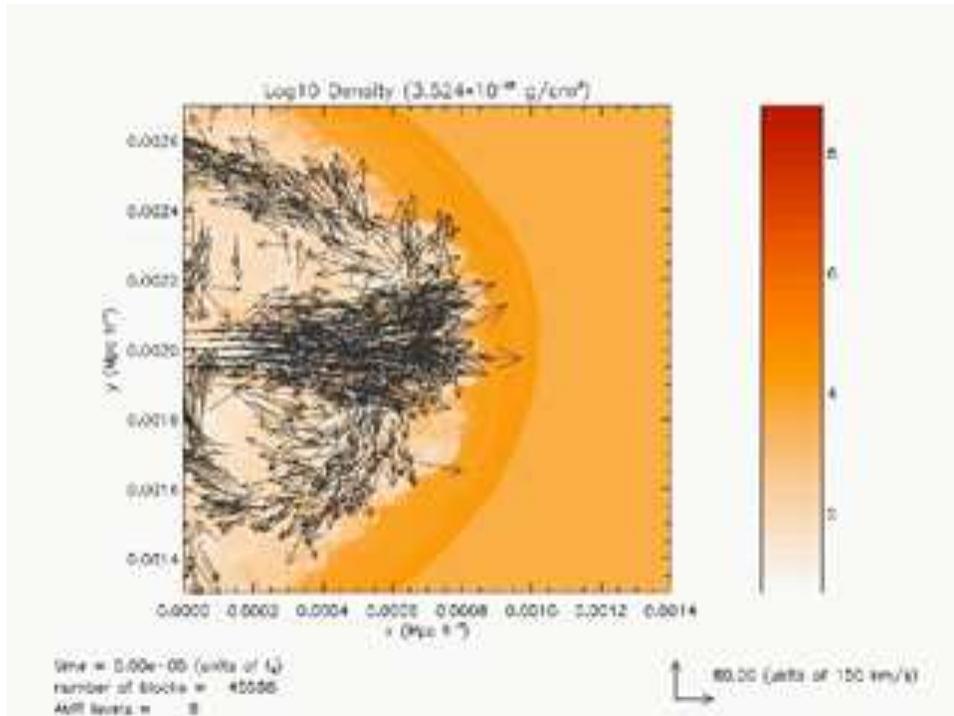, width=5truein}}}
\caption{\it 
Simulation of radio jet injection of energy into the ISM.  The jet is supersonic  and up to half the injected energy is dissipated by
driving turbulence  into an expanding overpressured coccoon.
Density and velocity distributions are shown.  
The high-density enhancement 
  within the cocoon is the stripped material from a cloud which has been shocked by the jet. The large-scale circulation in the cocoon 
  induces a backwards compression flow onto the protogalaxy
(Antonuccio and Silk 2007).}
\end{figure}

Some direct evidence for jet-induced star formation at $z>3$ comes from high redshift radio galaxy studies (Klamer et al. 2004, 2007),
where one sees the molecular gas 
reservoir as well as the young star distribution aligned with the radio lobes.
Minkowski's Object is a classical nearby example of jet-induced star formation
(Croft et al. 2006). More circumstantially, there is expected to be  a connection between AGN and ultraluminous  starbursts. ULIRGs are usually associated with a massive galaxy merger, also  favored as  supporting the case for the  monolithic gas conversion  phase evidenced by the trend in  the abundance ratios of alpha elements to iron with spheroid mass. Simulations suggest that  mergers  both fuel supermassive black hole growth and account for formation of early-type galaxies. 
Simulations suggest that the massive merger rate at $z\sim 4$ may suffice, up to one per massive halo, to account for massive spheroid formation (Fakhouri and Ma 2007).
Observational data on merger rates are consistent with the expectation from the simulations that the merger rate per halo per unit time increases approximately as $(1+z)^2.$
  \begin{figure}
\centerline{\hbox{
\epsfig{file=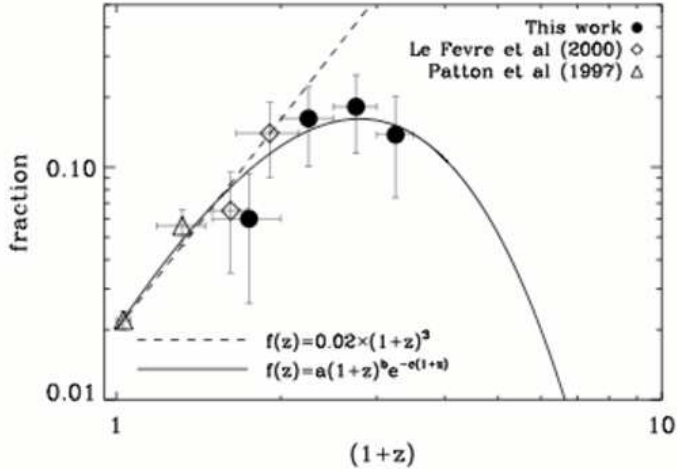, width=3.8truein}}}
\caption{\it 
Observed major merger fraction for massive galaxies ($\simgt 10^{10}\rm M_\odot$) as function of redshift.  Data  from Patton et al. (1997) and  Le Fevre et al. (2000), with added points
(filled circles) from Ryan et al. (2007).  The major  merger fraction  peaks at $z=1.8\pm 0.5$. Major merger activity coincides with the observed peaks in star formation history and AGN feeding.
 }
\end{figure}

However it has been  argued that 
major mergers do not seem to be  sufficiently frequent at $z\sim 2$ to account for the observed frequency of ULIRGs (Daddi et al. 2007).   This critique is partly based on the result from the Millennium
simulations that major mergers are predominantly dry by this epoch, as a consequence of the AGN feedback modelling recipe that is adopted (Springel et al. 2005).  One solution is appeal to 
minor mergers which are sufficiently common, to fuel generic spheroid formation,  along with providing a more extended star formation history. Another option is if AGN feedback is only effective at late epochs 
($z\simlt 2$) as inferred from observations  and modelling of cooling cores and
cavities in the intracluster gas (Nusser, Silk and Babul 2006; Best et al. 2007).

 Early major mergers would be gas-rich (wet), but late major mergers would then be 
gas-poor (dry). Semi-analytical modelling without AGN feedback indeed predicts that 
gas-rich (wet) mergers dominate in the past, and dry mergers are more abundant  at low redshift (Khochfar and Burkert 2001).The early mergers  could both feed the AGN 
and drive starbursts. Of course late epoch feedback, presumably by AGN, is still needed to account for 
the large current epoch  population of
red  galaxies.


%

\section{Where Next?}
 
The prospects are bright for advances in cosmology. Although all current 
observations are consistent with the hypothesis that $w=-1$ and constant,
new physics may be hiding within the error bars. The basic aim is to empirically 
track luminosity distance and perturbation growth factor over redshift to $z\sim 3.$ 
More than one technique will be essential,  with the aim of achieving better than 
 5\% 
accuracy in $w.$ In addition to supernovae, there are potentially three methods 
that appeal to different techniques and instrumentation: weak lensing, baryon oscillations 
and galaxy cluster surveys.

Another goal will be to find the first galaxies. Theory is not sufficiently precise to 
define the optimum search programmes: for example, it is not clear whether 
AGN precede, are contemporaneous with, or follow galaxy formation. Nor is 
dust formation and evolution understood, so that we cannot tell whether to focus on 
X-ray, optical, NIR, FIR or submillimetre frequencies.  At present, observers 
are undertaking or proposing  exploratory searches  over all of these frequency regimes. 
 
Current motivations center on developing a census of the earliest galaxies at $z=6$, 
or an age  of 0.95 Gyr.  With this in hand, one should be able to evaluate the
contribution of early star formation to cosmic reionisation, provided we understand 
the ionizing photon escape fraction. An important outcome will be improved 
constraints on mass assembly and feedback.

\subsection{ELTs and JWST}

In terms of probing the early universe, we are near the limits of all current facilities.  The 
planned restoration of HST in 2007 with installation of a new camera and 
spectrograph (WFC3, COS), and the repair of the existing camera and 
spectrograph (ACS, STIS) will provide a welcome boost to ultradeep survey 
cosmology. However the limitations of a 2.4m telescope are self-evident and  
now provide a major motivator for the ELTs and JWST. 

ELT design studies are focusing on AO-optimised 20-40m telescopes. The 
TMT (30m) is leading the pack at present with good progress on a \$80M design 
study (2004-2009) and a construction proposal now being considered by
various sponsors.  ESO has launched a 57M Euro design study for a 42m 
telescope(2007-2010). These projects envisage first light in 2016. 
AO will play a crucial role in ELT science and in particular multiconjugate AO
systems will be essential to give the fields essential for cosmology.
Exploring the territory with
laser-guide star systems on our existing telescopes is critical.

\subsection{21cm facilities}

The intergalactic medium at early epochs is the primary reservoir for galaxy formation,
and its structure will be probed by new generations of radio interferometers. An entirely
new domain will be opened with LOFAR and MWA via  21cm surveys for tomography of 
primordial HI. The Mileura Wide-Field Array will have a collecting area of 8000 m$^2$ over 1.5km
with arc-min resolution at 150MHz. In the era of TMT+JWST+LOFAR/MWA, we 
probably will not be interested in the simple issue of when reionisation occurred but rather 
how it occurred. Complex
physical processes involving varied astrophysical inputs will be studied and tracked by the topology and structure of ionization bubbles.
JWST can find luminous sources, and TMT will scan their  vicinities to determine 
topology of ionised shells via the distribution of fainter emitters. This will be  done in conjunction 
with HI surveys at the reionisation epoch.

\subsection{Theory}
  
Theory trails behind observation. It is reactive but lacks enough depth to be really predictive.
Our current understanding of
recent star formation is hierarchical, as viewed in disk galaxies. The star formation
history is extended, and there is evident for discrete gas feeding events at $z\sim 1.$
Most likely these are minor mergers. Cold gas feeding is required by theory in order 
to account for the continued instability of self-gravitating galactic disks to cloud and 
star formation. However the instability time-scale is a dynamical time, and one risks 
achieving far too high a star formation rate. This well-known problem is customarily resolved via feedback by supernova heating and momentum injection into the interstellar medium. This
accounts for the low fraction of gas converted into 
stars per dynamical time. However the physics is poorly modeled at current numerical resolutions. New approaches need to be developed to model feedback in multiphase interstellar media.

For massive spheroids,  the situation is at least superficially different. It is common to refer to star formation efficiency. Parenthetically, we recall that efficiency is an imprecise 
concept that combines time-scale and gas mass fraction converted into stars. To be more precise, we can specify star fraction formed per local dynamical time. Star formation was 
undoubtedly more efficient at high redshift than found today in disks. Downsizing occurred, with the most massive spheroids forming at even
higher efficiency. This seems to be a distinct starburst mode of star formation in which
the negative feedback by supernovae is unimportant. The starbursts  may be merger-triggered, but mergers also drive AGN feeding. Circumstantial evidence for AGN triggering, 
mostly in the past, suggests AGN outflows may also play a role in triggering the quasi-monolithic mode. What seems more certain is  that AGN outflows quench star formation by truncating the gas
supply. Here not only is current numerical resolution insufficient, but the relevant 
feedback physics is uncertain. From jets propagating in an inhomogeneous interstellar medium
to star formation is a major step that will require implementation of skills taken from two distinct communities of numerical simulators..

\section{Summary}

We require a healthy balance between general purpose facilities, projects targeting key 
science questions, and theory. Panoramic imaging with high resolution will 
revolutionise studies of dark matter on various scales. Multi-object and integral field spectroscopy
will enormously boost our understanding of galaxy evolution and provide far more reliable probes of dark energy.
 
A step-wise plan for constraining dark energy is emerging. Baryonic oscillation surveys
are the newest ingredient  and will generate much additional science.  Natural synergy 
is envisaged in studies of $z\sim 10$ galaxies between 30m telescopes with 
adaptive optics and mid-IR studies with JWST.  Laser guide star adaptive optics 
will revolutionise studies of  $z\sim 1-3$ emission line galaxies; ELTs will extend this to absorption line systems.
 
Massively parallel computing focussing on memory per node is essential to 
incorporate refined subgrid physics into the large-scale simulations.
Armed with improved star formation and AGN physics, we may eventually be able 
to develop robust predictions.  From these and the new generation of observations 
there will emerge answers to such key questions for the future as whether
hierarchical assembly of dark matter halos govern growth of galaxies. We need to ascertain
what physical processes curtail galaxy growth in a time-dependent manner, and what processes stimulate star formation. There is no doubt that 
exciting times are ahead with immense  challenges for observers and theorists alike.

\bigskip

\noindent{\bf References}

\noindent Abdalla, F. et al 2007 MNRAS in press (astro-ph/0705.1437)

\noindent Albrecht, A. et al Dark Energy Task Force report (astro-ph/0609951)

\noindent Allen, S. et al. 2006, MNRAS, 372, 21 

\noindent Antonuccio, V. and Silk, J. 2007, in preparation

\noindent Astier, P. et al 2006 A. \& A 447, 31 

\noindent
Best, P. et al. 2007,  MNRAS 379, 894

\noindent  Bouche, N. et al. 	 2007 preprint, arXiv0706.2656

\noindent Bouwens, R. et al 2007 Ap J in press,  arXiv:0707.2080

\noindent Bower, R.  et al. 2006  MNRAS, 370, 645

\noindent Croft, S., et al. 2006 Ap J 647, 1040 

\noindent  Croton, D.  et al. 2006 MNRAS 365, 11

\noindent  Daddi, E. et al. Ap J in press, 	arXiv:0705.2831

\noindent  Dekel, A.  \& Birnboim, Y. 2007 MNRAS in press, arXiv:0707.1214

\noindent Ellis, R.S. et al 2001 Ap J 560, L119

\noindent Ellis, R.S. 2007 {\it First Light in the Universe}, Saas Fee Lectures Series,
 Springer-Verlag, in press (astro-ph/0701024)

\noindent Ellis, R.S. et al 2007 Ap J in press (astro-ph/0710.3896)

\noindent Eyles, L. et al 2005 MNRAS 364, 443

\noindent Eyles, L. et al 2006 MNRAS 374, 910

\noindent Fakhouri, O. and  Ma, C. 2007, preprint, arXiv0710.4849

\noindent Fraternali, P. and Binney, J. 2007, preprint

\noindent Gavazzi, R. et al 2007 Ap J 667, 176

\noindent Glazebrook, K. et al 2007 in {\it Cosmic Frontiers}, eds. Metcalfe, N.
\& Shanks, T., in press, astro-ph/0701.876

\noindent Gottloeber, S. and Ypes, G. 2007
Ap J 664,  117

 \noindent Governato, F. et al.2007 MNRAS 374, 1479

\noindent Haiman, Z. et al 2005 astro-ph/0507013

\noindent Hasinger G., Miyaji T., and Schmidt M. 
2005, A\&A, 441, 417 

\noindent Heymans, C. et al 2006 MNRAS 368, 1323

\noindent Kashikawa, N. et al 2006 Ap J 648, L7

\noindent Kennicutt, R. 1998 Ap J 498, 541
	
\noindent Kennicutt, R. et al. 2007, preprint arXiv:0708.0922

\noindent Khochfar , S. \& Burkert, A. 2003 Ap J, 597, L117

\noindent Khochfar , S. \& Ostriker, J. 2007 Ap J submitted,  arXiv:0704.2418

\noindent
Klamer, I.  et al. 2004, Ap J, 612, L97

\noindent
Klamer, I.  et al. 2007 Ap J  in press, astro-ph/0703101 

\noindent Koopmans, L. et al 2006 Ap J 649, 599

\noindent Kneib, J-P. et al 2004 Ap J 607, 697

\noindent  Le Fevre, O. et al. 2000 MNRAS 311, 565

\noindent Massey, R. et al 2007a Nature, 445, 286

\noindent Massey, R. et al 2007b MNRAS 376, 13

\noindent Nesvada, N. et al 2007, A\&A in press, arXiv:0708.4150

\noindent Nagai, D. et al 2007 Ap J 655, 98

\noindent Navarro, J. et al 1997 Ap J 490, 493

\noindent Papovich, C. et al. 2006 Ap J 640,  92

\noindent  Patton et al. 1997 ApJ 475, 29

\noindent Peacock, J.A. et al 2007 ESA-ESO Report on Fundamental Cosmology (astro-ph/0610906)

\noindent Percival, W. et al 2007 MNRAS in press (astro-ph/0705.3323)

\noindent  Pope, A. et al. 2007 ApJ submitted  (astro-ph/0711.1153)

\noindent Ryan, R. et al. 2007, ApJ in press

\noindent Sajina, A. et al. 2007 ApJ 664, 713

\noindent Sand, D.J,. et al 2004 Ap J 604, 88

\noindent Sand, D.J. et al 2007 Ap J in press (astro-ph/0710.1069)

\noindent Semboloni, E. et al 2006 A \& A 452, 51

\noindent Schawinski, K.  et al. 2007
MNRAS in press,  arXiv:0709.3015 

\noindent Silverman, J. et al. 2007, preprint arXiv:0710.2461 

\noindent  Sommer-Larsen, J. 2006, ApJ 644, L1	

\noindent 
Silk, J.  and Rees, M. 1998, A\&A, 331, 1

\noindent  Springel. V. 2005, Nature 435, 629

\noindent Stark, D.P. et al 2007a Ap J 659, 84

\noindent Stark, D.P. et al 2007b Ap J 663, 10

\noindent Spergel, D. et al 2007 Ap J Suppl. 170, 377

\noindent  Thomas, D. et al. 2005 Ap J 621, 673

\noindent Tremonti, C. et al  2004, Ap J 613, 898

\noindent Yao, Y. et al  2007, ApJL in press,  arXiv:0711.3212

\end{document}